\documentclass[aps,prb,twocolumn,showpacs,superscriptaddress]{revtex4-1}

\usepackage{graphicx}
\usepackage{amssymb}
\usepackage{color}

\begin{document}

\title{Observation of crystallization slowdown in supercooled para-hydrogen and ortho-deuterium quantum liquid mixtures}

\author{Matthias K\"uhnel}
\affiliation{Institut f\"ur Kernphysik, J. W. Goethe-Universit\"at, Max-von-Laue-Str. 1, 60438 Frankfurt am Main, Germany}
\author{Jos\'e M. Fern\'andez}
\affiliation{Laboratory of Molecular Fluid Dynamics, Instituto de Estructura de la Materia, CSIC, Serrano 121, 28006, Madrid, Spain}
\author{Filippo Tramonto}
\affiliation{Laboratorio di Calcolo Parallelo e di Simulazioni di Materia Condensata, Dipartimento di Fisica, Universit\`a degli Studi di Milano, Via Celoria 16, 20133, Milano, Italy}
\author{Guzm\'an Tejeda}
\affiliation{Laboratory of Molecular Fluid Dynamics, Instituto de Estructura de la Materia, CSIC, Serrano 121, 28006, Madrid, Spain}
\author{Elena Moreno}
\affiliation{Laboratory of Molecular Fluid Dynamics, Instituto de Estructura de la Materia, CSIC, Serrano 121, 28006, Madrid, Spain}
\author{Anton Kalinin}
\affiliation{Institut f\"ur Kernphysik, J. W. Goethe-Universit\"at, Max-von-Laue-Str. 1, 60438 Frankfurt am Main, Germany}
\author{Marco Nava}
\affiliation{Laboratorio di Calcolo Parallelo e di Simulazioni di Materia Condensata, Dipartimento di Fisica, Universit\`a degli Studi di Milano, Via Celoria 16, 20133, Milano, Italy}
\affiliation{Computational Science, Department of Chemistry and Applied Biosciences, ETH Zurich, USI Campus, Via Giuseppe Buffi 13, CH-6900 Lugano, Switzerland}
\author{Davide E. Galli}
\affiliation{Laboratorio di Calcolo Parallelo e di Simulazioni di Materia Condensata, Dipartimento di Fisica, Universit\`a degli Studi di Milano, Via Celoria 16, 20133, Milano, Italy}
\author{Salvador Montero}
\affiliation{Laboratory of Molecular Fluid Dynamics, Instituto de Estructura de la Materia, CSIC, Serrano 121, 28006, Madrid, Spain}
\author{Robert E. Grisenti}\email[]{grisenti@atom.uni-frankfurt.de}
\affiliation{Institut f\"ur Kernphysik, J. W. Goethe-Universit\"at, Max-von-Laue-Str. 1, 60438 Frankfurt am Main, Germany}
\affiliation{GSI Helmholtzzentrum f\"ur Schwerionenforschung, Planckstr. 1, 64291 Darmstadt, Germany}


\begin{abstract}

We report a quantitative experimental study of the crystallization kinetics of supercooled quantum liquid mixtures of para-hydrogen (pH$_2$) and ortho-deuterium (oD$_2$) by high spatial resolution Raman spectroscopy of liquid microjets. We show that in a wide range of compositions the crystallization rate of the isotopic mixtures is significantly reduced with respect to that of the pure substances. To clarify this behavior we have performed path-integral simulations of the non-equilibrium pH$_2$-oD$_2$ liquid mixtures, revealing that differences in quantum delocalization between the two isotopic species translate into different effective particle sizes. Our results provide first experimental evidence for crystallization slowdown of quantum origin, offering a benchmark for theoretical studies of quantum behavior in supercooled liquids.

\end{abstract}

\pacs{64.70.dg, 02.70.Ss, 64.60.My, 67.63.Cd}



\maketitle

Understanding the stability of supercooled liquids with respect to crystallization is a fundamental open problem in condensed matter physics\cite{Ediger}. In this regard, since crystallization competes with glass formation, a knowledge of the mechanisms that govern the crystal growth in supercooled liquids is considered an important step to elucidate the nature of the glass transition\cite{Ediger2,Nascimento,Russo,Tang,Guerdane}. So far, experimental studies aiming at providing microscopic insights into dynamics and crystallization of supercooled liquids have been largely based on the use of colloidal suspensions\cite{Gasser,Hunter}, where the large particle size allows following the crystal growth on the laboratory time scale. However, diverse drawbacks such as polydispersity and sedimentation often make the experimental data from these systems difficult to interpret\cite{Hunter,Schoepe}. Accessing the details of the crystallization process in simple atomic and molecular counterparts, on the other hand, remains an experimental challenge due to relevant time scales that are orders of magnitude shorter.

Theoretical studies have shown that the inclusion of quantum effects adds a further degree of complexity in the behavior of supercooled liquids, leading to novel exotic phenomena like superfluidity\cite{Ginzburg,Osychenko} or enhanced dynamical slowing down\cite{Rabani,Markland,Markland2}. Yet again, the difficulties in supercooling a quantum liquid to very low temperatures have so far precluded possible experimental studies of the interplay of quantum effects and structural transformations in non-equilibrium bulk liquids. Here we address these challenges reporting on the experimental investigation of the crystallization kinetics of supercooled liquid mixtures of the isotopic species pH$_2$ and oD$_2$, showing that their quantum nature has a profound impact on the crystallization process.

Binary liquid mixtures exhibit in general properties that differ fundamentally from their corresponding pure substances, and mixing a few components is in particular a common strategy to hinder crystallization. Indeed, classical binary systems of particles that interact via a simple Lennard-Jones (LJ) pair potential have been widely employed as the simplest theoretical models to investigate crystallization and glassy behavior in supercooled liquids\cite{Coslovich,Valdes,Pedersen,Jungblut,Banerjee}. Due to the spherical symmetry of the ground-state wave function of the pH$_2$ and oD$_2$ molecules, which are characterized by an even rotational quantum number $J$\cite{Silvera}, a pH$_2$-oD$_2$ mixture provides a neat molecular binary system in which the pair interactions can be described by the same isotropic LJ potential\cite{Silvera}. Isotopic pH$_2$-oD$_2$ mixtures thus combine an intrinsic molecular simplicity with the exciting possibility to explore experimentally quantum behavior in supercooled liquids. It is in fact well established that the equilibrium thermodynamic and structural properties of the hydrogen liquids and solids are influenced by quantum effects\cite{Zoppi}. The magnitude of quantum effects can be quantified by the dimensionless parameter $\Lambda=\hbar/(\sigma\sqrt{m\epsilon})$, where $\hbar$ is the reduced Planck's constant; $\Lambda$ represents the (effective) de Broglie wavelength of a particle of mass $m$ relative to the length parameter $\sigma$ of the reference LJ potential characterized by a potential well depth $\epsilon$. For pH$_2$ and oD$_2$ one finds $\Lambda\approx 0.28$ and $\Lambda\approx 0.2$, respectively, which are one order of magnitude larger than the typical values for classical behavior. This quantum character of condensed pH$_2$ and oD$_2$ has led to the prediction of a variety of intriguing effects specific to the hydrogen liquids such as superfluidity of pH$_2$\cite{Ginzburg}, for which there is so far only indirect evidence coming from spectroscopic studies of small doped pH$_2$ clusters (see, e.g., Ref. \cite{Zeng} and references therein), or the realization of a structural quantum glass in a supercooled pH$_2$-oD$_2$ mixture\cite{Rabani}.

To reach a supercooled liquid state we have employed the experimental technique described in Ref. \cite{Kuehnel}. Briefly, the liquid at equilibrium pressure and temperature is injected into vacuum through a 5 $\mu$m-diameter glass capillary nozzle; the propagating liquid rapidly cools well below the melting temperature until it undergoes a first-order phase transition driven by the onset of homogeneous crystal nucleation, producing a continuous solid filament several cm long\cite{Kuehnel}. A crucial feature of our approach is represented by the univocal correspondence between the distance along the jet propagation direction, $z$, and time, $t=z/v$, where $v$ is the velocity of the liquid jet. We probed the crystallization kinetics of pH$_2$-oD$_2$ liquid jets with different oD$_2$ content via Raman light scattering by recording as a function of $z$ (and, thus, as a function of time) spectra of the fundamental vibrational transition, which allows distinguishing the liquid and solid phases of both the pH$_2$ and oD$_2$ components. The high spatial resolution of the present technique ultimately provides a direct access to the crystallization kinetics on the sub-microsecond time scale\cite{Kuehnel}. The isotopic species pH$_2$ and oD$_2$ were produced by continuous catalytic conversion from 99.9999\% and 99.9\% purity natural H$_2$ and D$_2$, respectively, resulting in 99.8\% and 97.5\% purity pH$_2$ and oD$_2$, respectively, the rest being represented by odd-$J$ molecules. We have investigated pure pH$_2$ and oD$_2$ liquid jets, as well as jets of pH$_2$-oD$_2$ mixtures with oD$_2$ mole percentages of $1\pm 0.02$, $3.2\pm 0.4$, $4.6\pm 0.6$, $9.1\pm 1.2$, $16.7\pm 0.8$, $51.6\pm 1.4$, $83.9\pm 0.7$, $95.7\pm 1.1$, and $97.6\pm 0.6$\%. The mixtures were prepared at room temperature by a continuous mixing of the two isotopic gases at the specific ratios set by two mass flow controllers, one for each species, working at a minimum flow rate of 20 nml min$^{-1}$. The oD$_2$ mole fraction has been further checked by the Raman intensity ratios in the gas (if available) and condensed phases in the vibrational region, confirming the mass flow ratios.

The experimental results are presented in Fig. 1. In Fig. 1(a) we show vibrational Raman spectra for three representative pH$_2$-oD$_2$ mixtures, clearly evidencing the liquid (L) to solid (S) phase transition. The observed shift to lower wave numbers of the vibrational band corresponding to the liquid with increasing distance from the orifice reflects the evaporative cooling of the propagating filament\cite{Kuehnel}. This is a consequence of the strong dependence of the vibrational wave number on temperature, though an explicit relation has been determined experimentally only for pure liquid pH$_2$\cite{Sliter}. In the case of our supercooled pH$_2$-oD$_2$ mixtures we have found that the vibrational wave number of liquid pH$_2$ at the onset of crystallization shows a linear dependence on the oD$_2$ mole fraction; accordingly, we have estimated the corresponding (average) temperature $T$ (shown by the upper $x$-axis of Fig. 2) by a linear interpolation between the experimental value $T\approx 12$ K for the pure pH$_2$ case\cite{Kuehnel} and $T\approx 17$ K as computed for pure oD$_2$ by a simple model that reliably describes the evaporative cooling of a liquid jet\cite{Kuehnel}.

In Fig. 1(b) we plot the time evolution of the pH$_2$ and oD$_2$ solid fractions extracted from the respective vibrational Raman spectra with the oD$_2$ mole fraction ranging from 0 (pure pH$_2$ jet, lower curve) to 1 (pure oD$_2$ jet, upper curve). The most striking feature is the remarkable slowdown of the crystallization kinetics with increasing oD$_2$. For example, the presence of only 3\% oD$_2$ molecules leads to nearly twice the time required for the complete freezing of the jet when compared to the pure pH$_2$ case ($\approx 7.9$ $\mu$s), as shown in Fig. 2. The slowest crystal growth is observed in the case of the nearly equimolar mixture, which fully crystallizes in $\approx 23.3$ $\mu$s, i.e. three times more slowly than the pure pH$_2$ jet. By further increasing the amount of oD$_2$ the duration of the crystallization process then gradually decreases down to $\approx 12.6$ $\mu$s for the pure oD$_2$ jet. A second important feature exhibited by the experimental data of Fig. 1(b) is the tendency of the filament to start crystallizing at earlier times with increasing oD$_2$ mole fraction, up to the case of the nearly equimolar mixture. This effect appears at first sight to be in conflict with the subsequent slower crystal growth. However, we can rationalize this behavior in terms of a higher probability for nucleation triggered by purely statistical clustering of oD$_2$ molecules, which reside at a much deeper supercooling than the pH$_2$ molecules with respect to their own melting points. The fact that the beginning of crystallization in mixtures with a higher content of oD$_2$ slightly shifts again towards later times results from the increasingly higher temperature of the filament (see Fig. 2).

The observed dependence of the jet crystallization time on composition as displayed in Fig. 2 is surprising given the isotopic nature of the pH$_2$-oD$_2$ mixtures. Due to the negligibly small mixing enthalpy\cite{White}, which determines the departure of a real mixture from the ideal case, there is no experimental evidence for a phase separation in H$_2$-D$_2$ mixtures at equilibrium neither in the liquid nor in the solid\cite{White}; this rules out possible effects related to the presence of strongly partitioning species, as observed in supercooled binary metallic alloys\cite{Herlach}. A possible competition between different crystal structures\cite{Valdes,Banerjee} seems also unlikely as both pH$_2$ and oD$_2$ crystallize into equilibrium hcp crystals\cite{Silvera}.
\begin{figure*}[t]
\includegraphics*[width=0.7\linewidth]{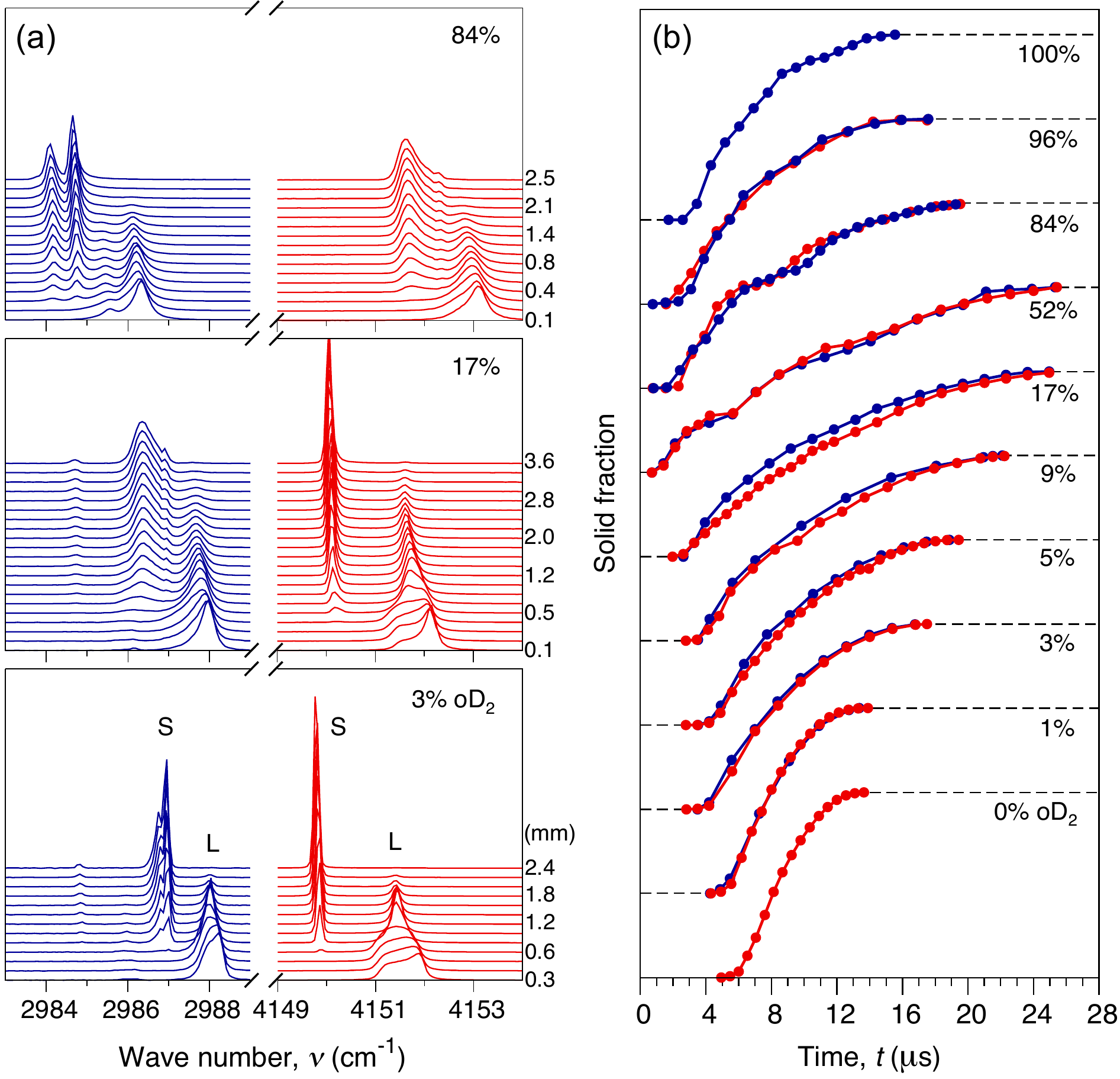}
\caption{\label{}(a) Normalized vibrational Raman spectra of the oD$_2$ (blue, left) and pH$_2$ (red, right) components measured as a function of the distance $z$ from the orifice (right scale) for three representative mixtures. The vibrational bands corresponding to the liquid and solid phases are indicated by L and S, respectively. The double-line shape of the oD$_2$ bands, which is especially evident in the case of the 84\% oD$_2$ mixture, is due to the inevitable presence of less than 3\% of $J=1$ pD$_2$ molecules, with a 50-fold enhancement in the Raman scattering intensity with respect to the $J=0$ molecules\cite{Kozioziemsky}. For the 3 and 17\% oD$_2$ mixtures this enhancement is much smaller and the $J=1$ bands are barely visible in the spectra. (b) Time evolution of the solid fractions extracted from the vibrational bands for pH$_2$ (red points) and oD$_2$ (blue points), with the time axis defined as $t=z/v$, where $v$ is the jet velocity. The solid fractions range from 0 to 1, as indicated by the dashed lines on the left and on the right of the experimental curves, respectively.}
\end{figure*}

The classical theory of crystal growth\cite{Jackson} is the natural framework to try interpreting our experimental crystallization rates. The crystal growth rate as a function of the temperature is given by $u(T) = k(T)\{1-{\rm exp}[-\Delta G(T)/k_BT]\}$, where $k(T)$ is the crystal deposition rate at the liquid/crystal interface, $\Delta G(T)$ is the difference in Gibbs free energy (per molecule) of the liquid and the crystal, and $k_B$ is Boltzmann's constant. For the case of our nearly ideal pH$_2$-oD$_2$ mixtures $\Delta G(T)$ can be computed starting from the experimental heat capacity data for the pure pH$_2$ and oD$_2$ systems\cite{Kuehnel}; we find that, for a given temperature, the factor $1-\exp\left[-\Delta G(T)/k_BT\right]$ varies only slightly with the amount of oD$_2$ or pH$_2$. Thus, our data suggest that the deposition rate $k(T)$ must be strongly dependent on composition in order to explain the measured crystallization rates. However, identifying this dependence is a challenging task since the coefficient $k(T)$, which reflects collective processes in the liquid, is generally expressed on an empirical basis\cite{Ediger2,Nascimento,Jackson}. We have recently shown that the crystallization of a pure pH$_2$ filament can be described by the collision-limited model\cite{Kuehnel}, in which the deposition rate scales as $k(T)\propto \sqrt{T/m}$. This mass dependence fairly captures the difference in the crystallization rates observed for the pure oD$_2$ and pH$_2$ jets; the relative duration of the crystallization processes in the two cases would be roughly given by the square root of the deuterium to hydrogen mass ratio, $\approx 1.4$, which is consistent with the experimental value of $\approx 1.6$ (Fig. 2). However, the above kinetic model is not able to describe the observed dependence of the crystal growth rate on composition, as in this case one would rather expect a monotonic increase of the jet crystallization time with oD$_2$ mole fraction.

As the next step we looked at the bulk structural features, which offer important insights into the behavior of supercooled liquids\cite{Russo,Coslovich,Pedersen,Leocmach}. To access static structural properties of the non-equilibrium pH$_2$-oD$_2$ quantum liquid mixtures we have carried out path-integral Monte Carlo (PIMC) simulations\cite{Ceperley} by using a canonical\cite{Rossi} Worm algorithm\cite{Boninsegni}. We have simulated mixtures of up to 300 molecules in boxes with periodic boundary conditions with 0, 3 and 10\% oD$_2$ at $T = 13$ K, with 50\% oD$_2$ at $T = 14.5$ K, and with 90, 97 and 100\% oD$_2$ at $T = 17$ K. In order to avoid the crystallization of the mixtures we followed the strategy reported in Ref. \cite{Osychenko}, verifying that the particles in the simulation cell remained in a disordered metastable configuration for a sufficiently large number ($\sim 10^4$) of Monte Carlo steps to ensure a reliable characterization of their physical properties. We have also verified that these latter were dependent only on the temperature and the density of the simulated system but not on the particular Monte Carlo stochastic trajectory sampled by checking that the results obtained from statistically independent simulations were consistent with each other.
\begin{figure}[t]
\includegraphics[width=0.7\linewidth]{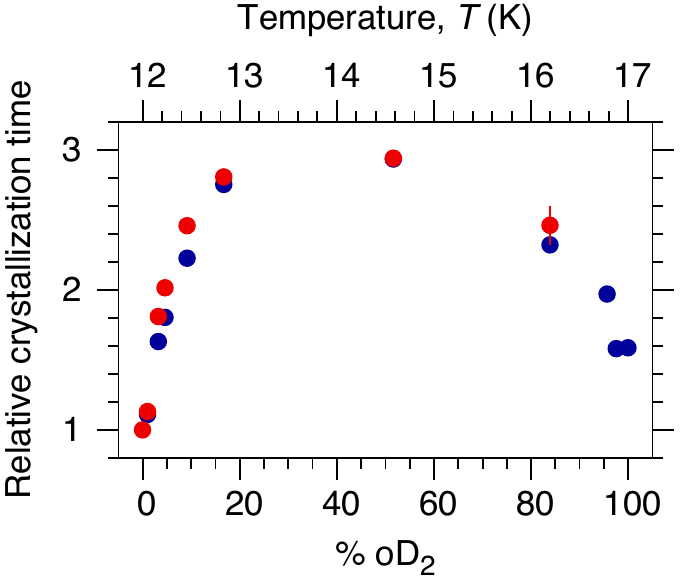}
\caption{Total duration of the jet crystallization process as determined from the pH$_2$ (red symbols) and oD$_2$ (blue symbols) solid fraction curves of Fig. 1(b). The plotted data indicate relative times with respect to the pure pH$_2$ jet crystallization time of $7.9\pm 0.4$ $\mu$s. The upper $x$-axis represents the estimated average filament temperature at the onset of crystallization as explained in the text.}
\end{figure}

In Fig. 3(a) we show a snapshot of the classical ring-polymers onto which the quantum particles are mapped computed for the 10\% oD$_2$ mixture. The degree of spatial extension of the polymers, each representing a pH$_2$ (red) or oD$_2$ (blue) molecule, is representative of the their quantum delocalization; Fig. 3(a) shows that the polymers associated to the oD$_2$ molecules are more compact than those associated to the pH$_2$ molecules. This feature affects the local structural properties, as shown in Fig. 3(b), where we plot the three partial radial pair distribution functions for the average static correlations. We see that the oD$_2$-oD$_2$ correlation exhibits a higher first peak, which is also shifted towards smaller distances than for the pH$_2$-pH$_2$ and pH$_2$-oD$_2$ correlations, indicating significant differences in the average distance between neighboring particles of the two isotopic species. Similar results are found for all the simulated mixtures. We point out that in a classical binary system the assumption of identical pair interactions would lead to identical static correlations for any relative fraction of the two isotopic species\cite{Lima}.
\begin{figure}[t]
\includegraphics[width=0.75\linewidth]{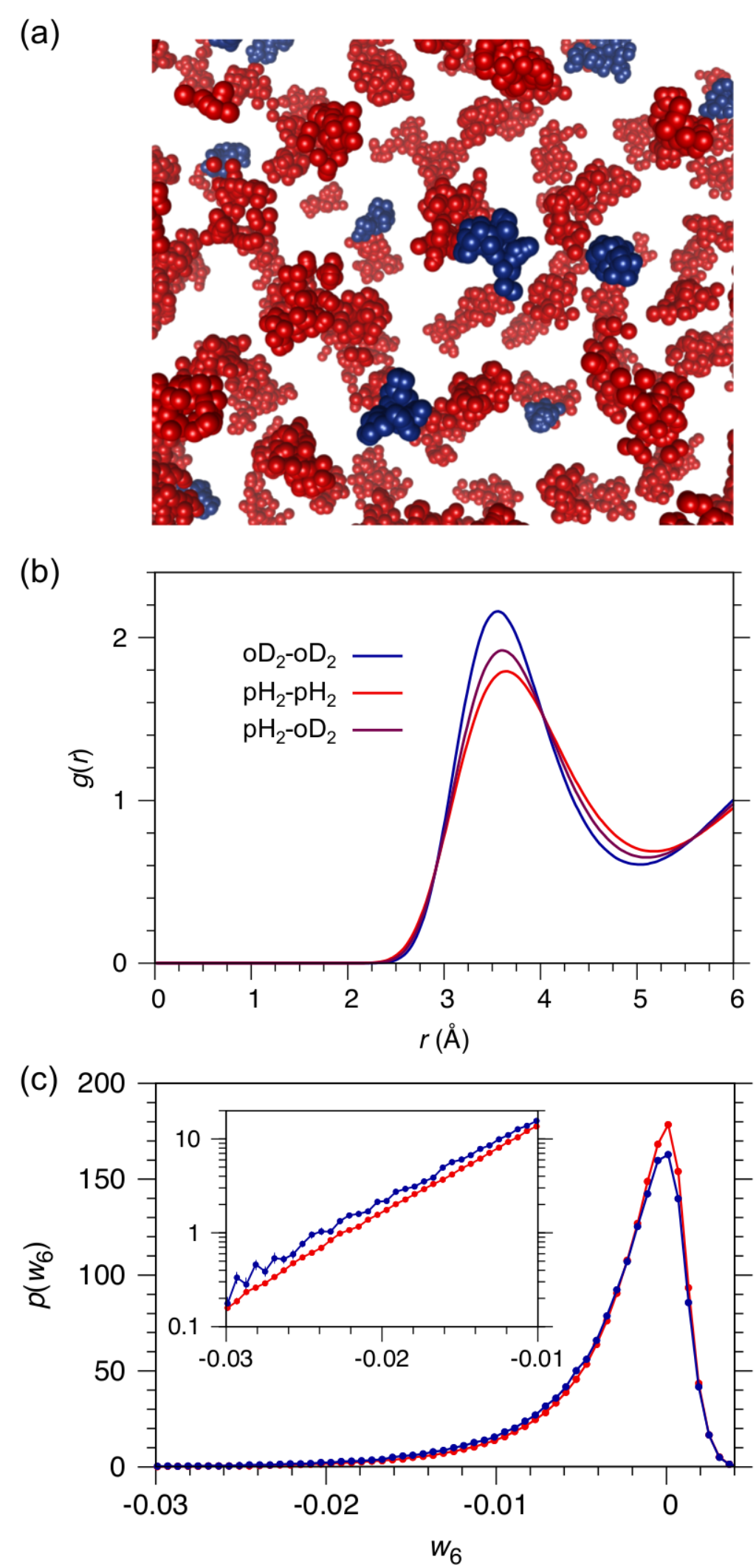}
\caption{\label{}PIMC simulations results for a mixture with 10\% oD$_2$ computed at $T = 13$ K. (a) Example of a ring-polymer configuration snapshot. The red and blue clusters of particles are the (classical) ring-polymers representing the pH$_2$ and oD$_2$ molecules, respectively. (b) Partial radial distribution functions $g(r)$ representing the three pair correlations. (c) Probability distribution $p(w_6)$ for icosahedral-like order for the pH$_2$ (red) and oD$_2$ (blue) molecules. The inset shows an enlarged view of the tail on a logarithmic scale, emphasizing the marked difference between the two isotopic species at high negative values of $w_6$.}
\end{figure}

The PIMC simulations show that as a result of the mass-induced quantum delocalization the pH$_2$ and oD$_2$ molecules exhibit different "effective sizes``, thus illuminating the origin of our measured growth rates. Classical molecular dynamics simulations of binary systems of particles with a given size ratio\cite{Valdes,Banerjee,Williams} have indeed reported a correlation between composition and crystallization kinetics that is strikingly similar to that observed in our experiments. In particular, for binary hard-sphere mixtures it was shown that the crystal growth becomes extremely slow for mole fractions of one of the two components in the range 20$-$50\% \cite{Williams}. A similar result was found in the simulation of a model binary LJ system\cite{Valdes}, evidencing in particular the failure of the mixture to crystallize, i.e. the formation of an amorphous state, for mole fractions of the smaller particles of 20$-$50\%, whereas rapid ordering has been observed otherwise. The similarity between those numerical results and our experimental data, which exhibit a maximum in the jet crystallization time for oD$_2$ mole fractions in the range 20$-$50\% (Fig. 2), is suggestive, hinting at a common mechanism responsible for the crystallization slowdown. However, a microscopic understanding of how composition and particle size ratio frustrate the crystal growth in binary mixtures is still lacking\cite{Williams,Valdes,Banerjee}.

A number of simulations studies\cite{Coslovich,Pedersen,Jungblut,Leocmach} have suggested that in one possible scenario crystallization might be hindered by the emergence in the bulk supercooled liquid of locally preferred structures that eventually are incompatible with long-range crystalline order\cite{Tarjus}. One important example is the icosahedron with its five-fold symmetry\cite{Frank}, which has been recently found to be a fundamental geometrical motif in the structure of bulk metallic glasses\cite{Hirata}. To explore this point we have performed a microscopic structural analysis based on the local bond order invariants method\cite{Russo,Jungblut,Leocmach} of our simulated pH$_2$-oD$_2$ mixtures. We have focused here on the invariant $w_6$, which is most sensitive to icosahedral-like order, but the present analysis does not rule out the presence in the metastable liquid of local order with different symmetries. In Fig. 3(c) we plot the probability distribution $p(w_6)$ computed separately for the pH$_2$ and oD$_2$ molecules for the 10\% oD$_2$ mixture. We see a slightly larger tendency for oD$_2$ to populate more negative values of $w_6$ than for pH$_2$, denoting an enhanced probability for local non-crystalline order around an oD$_2$ molecule. Similar results are found for all other simulated mixtures. As for the case of classical systems\cite{Leocmach}, we find that this tendency to non-crystalline order increases with the degree of supercooling. If a correlation between local order and crystal deposition rate could be established, then a difference in the packing efficiency for the two isotopic species as revealed by our structural analysis might provide a physical basis to explain the observed dependence of the crystal growth rate on the oD$_2$ mole fraction; the rearrangement of local non-crystalline structures in the supercooled melt at the liquid/crystal interface would tend to lower the particle diffusivity, thus slowing down the crystal growth\cite{Guerdane}.

Our experimental results show that composition and particle size ratio play a central role in the kinetics of crystallization of simple nearly-ideal molecular binary mixtures. Understanding the details of this dependence represents the key challenge that should allow identifying the principles that govern the stability of supercooled liquids against crystallization. With the present work we have not only established a promising experimental route to address this latter fundamental issue, but most notably we have provided first evidence for slowdown of crystallization that is purely of quantum origin. In fact, in the supercooled liquid pH$_2$-oD$_2$ mixtures it is only the difference in the mass-induced quantum delocalization of the two isotopic species that ultimately introduces a degree of frustration of crystallization.

We acknowledge financial support by the Deutsche Forschungsgemeinschaft, through Grant No. 593962, and the Spanish Ministerio de Ciencia e Innovaci\'on, through Grants No. FIS2010-22064-C02 and HD2008-0068. We acknowledge CINECA and the Regione Lombardia award, under the LISA initiative, for the availability of high performance computing resources and support. We acknowledge F. Gamez and D. Pini for useful discussions.

\newpage
%
%
%

%
%
%
%

\end{document}